\documentclass[prl,aps,twocolumn,superscriptaddress,nofootinbib,notitlepage,noeprint,notitle,nolongbibliography]{revtex4-2}
\usepackage{amsmath}
\usepackage{mathrsfs} 
\usepackage{amsbsy} 
\usepackage{bm}
\usepackage{xcolor}
\usepackage{ulem}
\usepackage{pdfsync}
\usepackage{graphicx}
\usepackage{amssymb}
\usepackage{cancel}
\usepackage{hyperref}
\usepackage[caption=false]{subfig}
\hypersetup{
    colorlinks,
    linkcolor={red!50!black},
    citecolor={blue!50!black},
    urlcolor={blue!80!black}
}

\usepackage{multirow}
\usepackage{xspace}

\usepackage{dcolumn}     
\newcolumntype{L}[1]{D{.}{.}{#1}}

\def\beq{\begin{align}}
\def\eeq{\end{align}}
\def\bea{\begin{eqnarray}}
\def\eea{\end{eqnarray}}
\newcommand{\bra}{{\langle}}
\newcommand{\ket}{{\rangle}}

\newcommand{\bs}[1]{\ensuremath{\boldsymbol{#1}}}
\def\nopi{$\cancel{\pi}$EFT\xspace}
\renewcommand{\vec}[1]{\ensuremath{\boldsymbol{#1}}}

\begin{document}

\title{Nuclear Structure Effects on Hyperfine Splittings in Ordinary and Muonic Deuterium}

\author{Chen Ji}
\email{jichen@ccnu.edu.cn}
\affiliation{Key Laboratory of Quark and Lepton Physics, Institute of Particle Physics, Central China Normal University, Wuhan 430079, China}
\affiliation{Southern Center for Nuclear-Science Theory, Institute of Modern Physics, Chinese Academy of Sciences, Huizhou 516000, China}

\author{Xiang Zhang}
\affiliation{Key Laboratory of Quark and Lepton Physics, Institute of Particle Physics, Central China Normal University, Wuhan 430079, China}

\author{Lucas Platter}
\affiliation{Department of Physics and Astronomy, University of
Tennessee, Knoxville, TN 37996, USA}
\affiliation{Physics Division, Oak Ridge National Laboratory, Oak Ridge, TN 37831, USA}


\date{\today}

\begin{abstract}
  Precision spectroscopy of hyperfine splitting (HFS) is a crucial
  tool for investigating the structure of nuclei and testing quantum
  electrodynamics (QED). However, accurate theoretical predictions are
  hindered by two-photon exchange (TPE) effects. We propose a novel
  formalism that accounts for nuclear excitations and recoil in TPE,
  providing a model-independent description of TPE effects on HFS in
  light ordinary and muonic atoms. Combining our formalism with
  pionless effective field theory at next-to-next-to-leading order,
  the predicted TPE effects on HFS are 41.2(2.6) kHz and 0.116(9) meV
  for the 1S state in deuterium and the 2S state in muonic
  deuterium. These results are within 1.4-1.7$\sigma$ from recent
  measurements and highlight the importance of nuclear structure
  effects on HFS and indicate the value of more precise measurements
  in future experiments.
\end{abstract}

\maketitle

\paragraph*{Introduction.}
\label{sec:intro}
Precision laser spectroscopy of atomic transitions informs on the
structure of nuclei and tests the accuracy of bound-state quantum
electrodynamics (QED). Measurements of Lamb shifts in light, muonic
atoms have provided nuclear charge radii at unprecedented
accuracy~\cite{Pohl:2010zza,Antognini13,Pohl:2016sc,CREMA:2023blf,Krauth2021}. In these
experiments, a solid understanding of nuclear structure effects is crucial~\cite{Pachucki:2011xr,Ji13,Ji:2018ozm,Nevo_Dinur_2016,Hernandez2018}.

High-precision spectroscopy measurements of hyperfine splitting (HFS)
have provided valuable insights into the nuclear magnetic
structure. These measurements have been conducted on light atoms such
as $^{1,2}$H, $^3$He, and
$^{6,7}$Li~\cite{Hellwig:1970,Wineland:1972pra,Rosner:1970pra,Kowalski:1983hi,Guan:2020pra},
as well as their muonic counterparts including $\mu^{1,2}$H and
$\mu^3$He$^+$~\cite{Antognini13,Pohl:2016sc,CREMA:2023blf}. HFS,
predominantly governed by the short-range interaction between the
nuclear and lepton magnetic
moments~\cite{Schwartz:1955pr,Woodgate:1983de,Eides2001}, offers an
ideal probe for studying the elastic and inelastic structure of
nucleons and nuclei. 

Accurate theoretical predictions for HFS
in both ordinary and muonic atoms are limited by nuclear structure
effects, entering through two-photon exchange (TPE). The elastic TPE, encoded in the \textit{Zemach radius} $r_Z$, arises
from the convolution of the nuclear charge and magnetic
densities~\cite{Zemach:1956zz,Friar:2004plb}. The inelastic TPE,
namely the nuclear polarizability, stems from nuclear virtual excitations.

For $^2$H and $\mu^2$H, the discrepancy between the measured HFS and
the calculated QED contribution for the 1S state of $^2$H
is~\cite{Wineland:1972pra,Eides2001}.
\begin{align}
\label{eq:tpe-D-exp}
\nu_{\rm exp}(^2\text{H})-\nu_{\rm QED}(^2\text{H}) = 45.2 \;\text{kHz},
\end{align}
and for the 2S state of $\mu^2$H is~\cite{Krauth:2015nja,Kalinowski:2018}
\begin{align}
  \label{eq:tpe-muD-exp}
 \nu_{\rm exp}(\mu^2\text{H})-\nu_{\rm QED}(\mu^2\text{H}) = 0.0966(73)\;\text{meV}.
\end{align} 
These discrepancies mainly arise from TPE. However,
an accurate, uncertainty-quantified, and model-independent prediction of the TPE
effect on HFS has not been achieved
yet~\cite{Faustov:2003pra,Friar:2005yv,Friar:2005rc,Khriplovich:1995yd,Khriplovich:2003nd,Kalinowski:2018}. For
instance, the conventional Low-term formalism inadequately accounts for
nuclear excitations, thus providing an incomplete
description~\cite{Friar:2005yv,Friar:2005rc}.

This paper introduces a new formalism for the TPE effect on HFS, that
accurately incorporates nuclear excitations and recoil. Using pionless
effective field theory (\nopi) at next-to-next-to-leading order
(NNLO), we then evaluate TPE contributions in $^2$H and $\mu^2$H. The
formalism offers a model-independent description of the TPE effect
with systematic uncertainty quantification, showing consistency with
$\nu_{\rm exp}$ - $\nu_{\rm QED}$ in $^2$H and $\mu^2$H.

\paragraph*{Two-Photon Exchange Theory.}

HFS of $ns_{1/2}$ states is dominated by contact interactions between
the lepton spin $\vec{\sigma}_\ell/2$ and the nuclear spin
$\vec{I}$~\cite{Schwartz:1955pr,Woodgate:1983de,Eides2001}
\begin{equation}
\mathcal{H}_I = \frac{2\pi\alpha g_m }{3 m_\ell m_N} 
\phi^2_n(0) 
	\vec{\sigma}^{(\ell)} \cdot \vec{I},
\end{equation}
where $\alpha$ is the electromagnetic fine structure constant, $g_m$
denotes the nuclear magnetic g-factor, and $m_\ell$ ($m_N$) is the
lepton (nucleon) mass. $\phi_n^2(0)=(Z\alpha)^3 m_R^3/(n^3\pi)$ is the
wave function squared of the atomic $ns_{1/2}$ state at the origin,
with $m_R$ denoting the lepton-nucleus reduced mass.  Its contribution
to HFS is at $\alpha^4$ and is evaluated as the expectation on the
atomic hyperfine state by
\begin{align}
\label{eq:HFS-LO}
  E_F = \bra (ns_{1/2}, N_0I) F M_F| \mathcal{H}_I | (ns_{1/2}, N_0I) F M_F\ket,
\end{align}
where $|N_0 I\ket$ is the nuclear ground state with spin $I$, and $F$
and $M_F$ denote the total angular momentum and its z-projection.

\begin{figure}[t]
\centering
\includegraphics[width=0.9\linewidth]{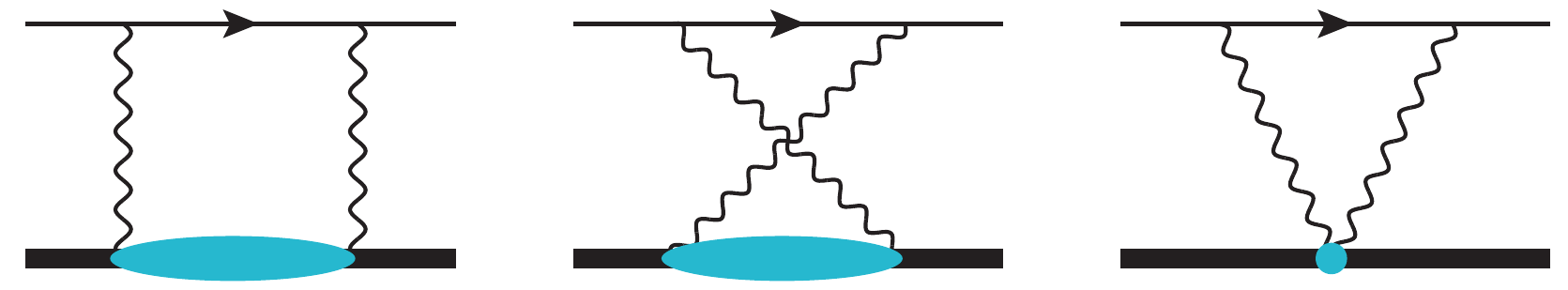}
 \caption{Doubly virtual two-photon exchange diagrams.}
\label{fig:tpe}
\end{figure}

The TPE effect arises at $\alpha^5$, driven by doubly virtual photon
exchanges between the nucleus and the lepton, as illustrated in
Fig.~\ref{fig:tpe}. The corresponding operator is expressed in Lorenz gauge as~\cite{Friar:2005rc}
\begin{align}
\label{eq:H-2g}
\mathcal{H}_{\rm 2\gamma} = 
i (4\pi\alpha)^2 
\phi_n^2(0)
\int \frac{d^4 q}{(2\pi)^4}
\frac{\eta_{\mu\nu}(q)\, T^{\mu \nu}(q,-q) }{(q^2+i\epsilon)^2(q^2-2m_{\ell} q_0+i\epsilon )}~,
\end{align}
where $\eta$ and $T$ respectively represent the lepton and nuclear
tensors. Only the lepton-spin
dependent part
$\tilde{\eta}^{\mu \nu} = i q_0 \epsilon^{0 \mu \nu
  i}\sigma_i^{(\ell)} + i \epsilon^{\mu \nu i j} \sigma^{(\ell)}_i
q_j$ of the lepton tensor contributes to HFS. The third diagram in
Fig.~\ref{fig:tpe} is the nuclear seagull tensor $B_{\mu \nu}$. The
charge-current part $B_{0m}$ is of relativistic order at
$1/m_N^2$. The current-current part $B_{ij}$ gets canceled due to
crossing symmetry~\cite{Friar:2005rc,Friar:1974}.

\paragraph*{TPE polarizability.}
We find the inelastic TPE operators by using the spin-dependent part of the lepton
tensor and incorporating a summation over nuclear excitations in the
nuclear tensor~\cite{Friar:2005rc}
\begin{align}
\mathcal{H}_{\rm pol}^{(0)} =&  \frac{ i \alpha^2 \phi_n^2(0)}{2 \pi m_{\ell}^2 } 
\int d\omega \int \frac{d^3 q }{\vec{q}^4} h^{(0)}(\omega, |\vec{q}|)
                               \nonumber
\\
&\vec{\sigma}^{(\ell)} \cdot \left\{ \vec{q} \times \vec{J}(-\vec{q}), J_0(\vec{q}) \right\} \delta(\omega-\omega_N)~,
\\
\mathcal{H}_{\rm pol}^{(1)} =& \frac{ i \alpha^2 \phi_n^2(0)}{2 \pi m_{\ell}^2}
\int d\omega \int \frac{d^3 q }{\vec{q}^2} h^{(1)}(\omega, |\vec{q}|)
\nonumber\\
&\vec{\sigma}^{(\ell)} \cdot \left[\vec{J}(-\vec{q}) \times \vec{J}(\vec{q}) \right] \delta(\omega-\omega_N)~,
\end{align}
where $\omega_N$ denotes the excitation energy of the nuclear
state. $\mathcal{H}_{\rm 2\gamma}^{(0)}$ involves the charge-current
transition matrix with the two operators in
anti-commutation. $\mathcal{H}_{2\gamma}^{(1)}$ involves the
current-current matrix with the two currents in commutation, and is
one order higher in $1/m_N$. The kernels $h^{(0,1)}$ are
\begin{align}
  h^{(0}(\omega,q) =& \left[2+\frac{\omega}{E_q}\right]
\frac{E_q^2+m_{\ell}^2+E_q \omega}{(E_q+\omega)^2-m_\ell^2 }   - \frac{2q+\omega}{q+\omega}~,
  \\
    h^{(1)}(\omega,q) =& \frac{1}{E_q}  
                         \frac{E_q^2+m_{\ell}^2+E_q \omega}{(E_q+\omega)^2-m_\ell^2 }  - \frac{1}{q+\omega}~,
\end{align}
with $E_q =\sqrt{q^2+m_{\ell}^2}$.

To obtain the polarizability corrections $E_{\rm pol}^{(0,1)}$, we
replace $\mathcal{H}_{\rm I}$ with $\mathcal{H}_{\rm pol}^{(0,1)}$ in
Eq.~\eqref{eq:HFS-LO}. Using the Wigner-Eckart theorem, we factorize
the lepton and nuclear matrix elements in $E_{\rm pol}^{(0,1)}$,
expressing them as ratios to $E_F$. We write $J_0$ as charge density
$\rho$ and decompose $\vec{J}$ into convection ($\vec{J}_c$) and
magnetic ($\vec{J}_m$) currents. This leads to photo-induced nuclear
sum rules:
\begin{align}
\label{eq:Epol0}
E_{\rm pol}^{(0)} 
=&\frac{ 6\alpha m_N E_F}{\pi m_l g_m I}   
\int_{\omega_{\rm th}}^\infty d\omega \int_0^\infty d q h^{(0)}(\omega, q) S^{(0)}(\omega,q), 
\\
\label{eq:Epol1}
E_{\rm pol}^{(1)} 
=& - \frac{ 6\alpha m_N E_F }{\pi m_l g_m I}  
\int_{\omega_{\rm th}}^\infty d\omega \int_0^\infty d q
 h^{(1)}(\omega, q) S^{(1)}(\omega,q),
\end{align}
where $\omega_{\rm th}=(\gamma^2+q^2/4)/m_N$ is the minimum deuteron
excitation energy in the inelastic TPE. The nuclear excitations in the
deuteron are represented by the scattering state $|\psi_{\bs
  p}\ket$. The deuteron charge-magnetic ($S^{(0)}$) and
convection-magnetic ($S^{(1)}$) response functions are
\begin{align}
\label{eq:S0}
S^{(0)}(\omega,q) 
=& \frac{m_N p}{64\pi^4 q^2} \iint d \hat{p}\, d \hat{q}
\nonumber \\
&\hspace{-12ex} 
{\rm Im} \left(
\bra N_0 I I | \rho(-\vec{q})  | \psi_{\vec{p}} \ket 
\bra  \psi_{\vec{p}} | \left[\vec{q} \times \vec{J}_m(\vec{q})\right]_3 | N_0 I I \ket 
\right)
\\
\label{eq:S1}
S^{(1)}(\omega,q) 
=& \frac{m_N p}{64\pi^4 } \iint d \hat{p}\, d \hat{q}
\epsilon^{3jk} 
\nonumber \\
&\hspace{-12ex}
\times{\rm Im} \left(
\bra N_0 I I | \vec{J}_{c,j}(-\vec{q}) | \psi_{\vec{p}} \ket 
\bra  \psi_{\vec{p}} | \vec{J}_{m,k}(\vec{q})  | N_0 I I \ket 
\right),
\end{align}
where $|N_0 I I \ket$ denotes the nuclear ground state with spin
maximally projected in the z-direction. For the deuteron, its nuclear
excitation involves the two-nucleon scattering states at relative momentum $p = \sqrt{m_N \omega - \gamma^2 - q^2/4}$.

\paragraph*{Elastic TPE.} The elastic TPE contribution involves the insertion of the momentum-boosted nuclear
ground state into the nuclear tensor $T_{\mu \nu}$, leading to
\begin{align}
\label{eq:el0}
E_{\rm el}^{(0)} 
=& \frac{ 2 \alpha E_F}{ \pi m_l }   
\int_0^\infty d q 
\left[h^{(0)}(\frac{q^2}{4m_N}, q) F_{md}(q) F_{ed}(q) 
\right.
\nonumber \\
& 
\left. 
- \frac{4 m_l m_R}{q^2} \right],
\\
\label{eq:el1}
E_{\rm el}^{(1)}
=& - \frac{ \alpha E_F}{2\pi m_l m_N}   
\int_0^\infty d q  q^2 
h^{(1)}(\frac{q^2}{4m_N}, q) F_{md} (q) 
F_{ed}(q),
\end{align}
where $q^2/(4m_N)$ is the deuteron recoil energy in the elastic TPE process.
The deuteron electric and magnetic form factors, $F_{ed}$ and $F_{md}$
are normalized to 1 at $q=0$. The function $h^{(0)}$ is approximated
by $4m_l m_R/q^2$ when taking $m_N \gg m_l$, changing
$E_{\rm el}^{(0)}$ to the pure Zemach contribution
$E_{\text{zem}} = -2\alpha m_R
r_Z$~\cite{Zemach:1956zz,Friar:2004plb}. The subtraction term in
Eq.~\eqref{eq:el0} cancels the infrared divergence of the
$q$-integration and prevents a double counting in the iteration of the
lowest-order single-photon exchange in the point-nucleus
limit~\cite{Carlson:2011pra}. $E_{\rm el}^{(1)}$ is also a convolution
of nuclear magnetic and electric densities but is suppressed by
$1/m_N$ relative to $E_{\rm el}^{(0)}$.
 
A higher-order correction to $E_{\rm el}^{(0)}$ arises from the deuteron S-to-D-state mixing, and is given by
\begin{align}
\label{eq:el2}
E_{\rm el-sd}^{(0)} = \frac{\alpha \mu_Q E_F}{3\pi m_l} \int^\infty_0 dq q^2 h^{(0)}(\frac{q^2}{4m_N}, q) F_{md}(q) F_{Qd}(q),
\end{align}
where $F_{Qd}$ denotes the deuteron quadrupole form factor, which is normalized to 1 at $q=0$.

\paragraph*{Single-nucleon TPE.} An additional correction to HFS arises from TPE
between the lepton and a single nucleon, and includes the nucleon's
Zemach, recoil, and polarizability effects. When embedded in a
nucleus, the single-nucleon TPE contributions in $^2$H and $\mu^2$H
are~\cite{Kalinowski:2018}
\begin{align}
E_{1N} = -\frac{2\alpha m_l E_F}{g_m(m_l+m_p)}   \left(\kappa_p  \tilde{r}_Z^p  + \kappa_n  \tilde{r}_Z^n  \right).
  \label{eq:rzN}
\end{align}
where $\tilde{r}_Z^p$ and $\tilde{r}_Z^n$ represent the effective
proton and neutron Zemach radii, accounting for the full
single-nucleon TPE effects
\cite{Tomalak:2019epja,Tomalak:2019prd,Antognini:2022arn,Hagelstein:2015lph,Hagelstein:2018bdi}. 


\paragraph*{Pionless effective field theory.}
\label{sec:nopi}

We employ the identical Lagrangian used in Ref.~\cite{Emmons:2020aov}
to compute the two-nucleon bound and scattering states utilizing
dimensional regularization and power-divergence subtraction (PDS)
renormalization. In addition, we include the NNLO S-to-D-wave mixing
operator~\cite{Chen:1999vd, Ji:2003ia, Ando:2004mm}
\begin{align}
\mathcal{L}_{s d}
=&\frac{C_0^{(s d)}}{4} d_i^\dagger\biggl[N^T P^{j} \bigl(\tensor{\nabla}_i \tensor{\nabla}_j - \frac{\delta_{ij}}{3}  \tensor{\nabla}^2 \bigr) N\biggr]
+ \text{h.c.}~,
\end{align}
where
$\tensor{\nabla} \equiv \overleftarrow{\nabla} -
\overrightarrow{\nabla}$, $\gamma$ denotes the deuteron binding
momentum and $\mu$ the PDS renormalization scale.
$C_0^{(sd)} = - 6\sqrt{2}\pi \eta_{sd} /[m_N \gamma^2 (\mu-\gamma)]$
~\cite{Chen:1999vd, Ji:2003ia, Ando:2004mm} matches the deuteron's
asymptotic D-to-S wave ratio $\eta_{sd}=0.0252$~\cite{Stoks:1994wp}.

$P$-wave contact interactions enter \nopi at
N$^3$LO~\cite{Rupak:2000npa, Chen:1999prc}. Furthermore, the
relativistic correction to the kinetic term is suppressed by
$1/m_N^2$~\cite{Chen:1999tn}, thus of N$^4$LO size. We neglect these
higher-order contributions in this work.

The one-nucleon current originates from minimal substitution in the
free part of the Lagrangian and is~\cite{Chen:1999tn, Rupak:2000npa}.
\begin{multline}
    \label{eq:Lem-1b}
\mathcal{L}_{\text {EM },1 b}=-\frac{e}{2} N^\dagger [F_{es}(q)+\tau_3 F_{ev}(q)] N A_0\\
  -\frac{i e}{4 m_N} N^\dagger \tensor{\nabla} [F_{es}(q)+\tau_3 F_{ev}(q)] N \cdot \vec{A}
  \\
+\frac{e}{2 m_N} N^\dagger\left[\kappa_0 F_{ms}(q)+\kappa_1 \tau_3 F_{mv}(q)\right] \vec{\sigma} \cdot \vec{B} N,
\end{multline}
where $\vec{\sigma}$ denotes the nucleon Pauli matrix vector. The
nucleon isoscalar and isovector anomalous magnetic factors denoted as
$\kappa_0$ and $\kappa_1$, are related to the magnetic factors of the
proton and neutron by $\kappa_0 = (\kappa_p + \kappa_n)/2$ and
$\kappa_1 = (\kappa_p - \kappa_n)/2$. For the nucleon electric and
magnetic isoscalar and isovector form factors, we use the
recent parametrization based on dispersion analysis of the time- and space-like $eN$ scattering data~\cite{Lin:2021umz,Lin:2021umk,Lin:2021xrc}.

Two-nucleon currents appear at higher orders in \nopi. Introducing
covariant derivatives in the $np$ spin-triplet interaction gives rise
to a two-nucleon convection current at NLO, whose interaction
Lagrangian is
\begin{equation}
\label{eq:L_2bc-c}
\mathcal{L}_{2,C} = \frac{i e C_2}{4} F_{ev}(q) d_i^\dagger (N^T \tensor{\nabla} P_i \tau_3 N)\cdot \vec{A} + \text{h.c.},
\end{equation}
where $C_2$ represents the known coefficient of the two-nucleon NLO
interaction~\cite{Chen:1999tn}. $\mathcal{L}_{2, C}$ does not
contribute to nuclear electric form factors but affects nuclear
polarization. Furthermore, the two-nucleon magnetic current, which
couples with the $np$ spin-triplet interaction, emerges at NLO but not
through minimal substitution:
\begin{align}
\label{eq:L-2bc-m}
 \mathcal{L}_{2, B}=
 - i e L_2 F_{ms}(q) \epsilon_{i j k}d_i^\dagger d_j B_{k} + \text{h.c.},
\end{align}
with $L_2 = (g_m - 2\kappa_0 )\pi/ [2 m_N\gamma(\mu-\gamma)^2]$
determined by matching to the measured magnetic g-factor. The
two-nucleon magnetic current causes the $np$ spin-singlet-to-triplet
transition to emerge at NLO but does not contribute to TPE in HFS due
to spin-parity selection rules. Other two-nucleon currents are beyond
NNLO~\cite{Chen:1999tn,Rupak:2000npa}.

The transition matrices necessary for calculating the response
functions in Eqs.~(\ref{eq:S0},\ref{eq:S1}) are determined using a
similar approach as in Ref.~\cite{Emmons:2020aov}. The detailed
expressions can be found in the supplementary material~\cite{supp2023}.

\begin{figure}[th]
\centering
\includegraphics[width=0.8\linewidth]{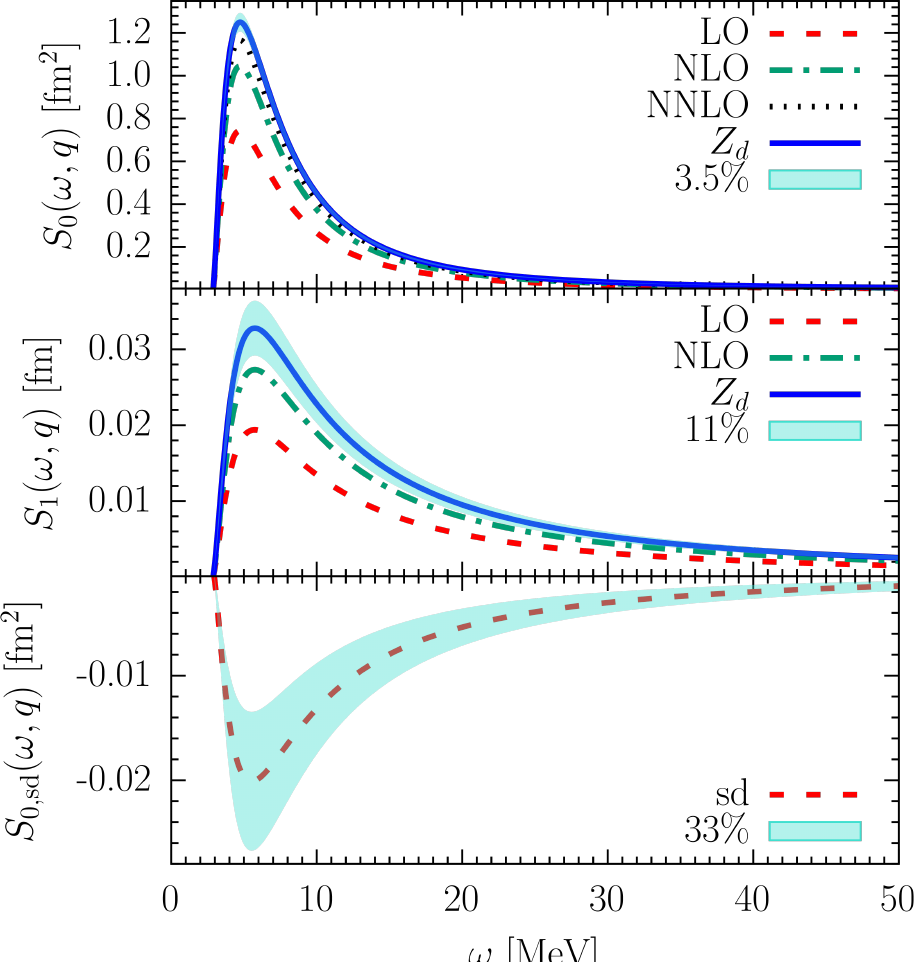}
\caption{
\label{fig:Sfull}
The response functions $S^{(0)}$ (top panel), $S^{(1)}$ (middle
panel), and $S^{(0)}_{\text{sd}}$ (bottom panel) are shown as
functions of $\omega$ for a fixed $q=50$ MeV. The leading,
sub-leading, sub-sub-leading, and $Z_d$-improved results are
represented by the red dashed, green dot-dashed, black dotted, and
blue solid lines, respectively (color online). The light-blue band
represents the uncertainty error from omitted higher-order
corrections.}
\end{figure}

\paragraph*{Results.}
The TPE correction to HFS consists of the elastic, polarizability, and single-nucleon contributions
\begin{align}
	E_{\rm TPE}^{\rm HFS} = E_{\rm el} + E_{\rm pol} + E_{1p} + E_{1n}.
\end{align}
The response functions in Eqs.~(\ref{eq:S0}, \ref{eq:S1}) are
numerically evaluated in the \nopi framework. The regulator
independent results indicate the model independence of the
prediction. Fig.~\ref{fig:Sfull} displays the charge-magnetic response
function $S^{(0)}$, S-D mixing correction $S^{(0)}_{\text{sd}}$, and
the convection-magnetic one $S^{(1)}$, as functions of the excitation
energy $\omega$ at a fixed transfer momentum $q = 50$ MeV. $S^{(0)}$,
which dominates in the polarizability effect, is calculated at NNLO,
while $S^{(1)}$, whose contribution is suppressed by $1/m_N$, is
evaluated at NLO. Following
Ref.~\cite{Emmons:2020aov,Phillips:1999hh,Phillips:1999am},
$S^{(0,1)}$ are $Z_d$-improved for better accuracy by accounting for
the remaining effective range correction in the deuteron asymptotic
normalization constant. A relative uncertainty of $3.5\%$ ($11\%$) is
assigned to $S^{(0)}$ ($S^{(1)}$), due to omitted N$^3$LO (NNLO)
corrections. $S^{(0)}_{\mathrm{sd}}$, expected to enter at NNLO,
carries a relative uncertainty of $33\%$ due to its omitted
higher-order correction. Inserting the response functions in
Eqs.~(\ref{eq:Epol0},\ref{eq:Epol1}) leads to the polarizability
effects
$E_{\rm pol}=E_{\text{pol}}^{(0)}+E_{\text{pol}}^{(1)}+E_{\rm
  pol,sd}^{(0)}$.

With the deuteron form factors evaluated in \nopi, the elastic TPE is
a summation of contributions in Eqs.~(\ref{eq:el0}-\ref{eq:el2}),
$E_{\rm el}=E_{\text{el}}^{(0)}+E_{\text{el}}^{(1)}+E_{\rm
  el,sd}^{(0)}$. It is different from the Zemach contribution
$E_{\text{zem}}$ due to the additional recoil corrections in
$E_{\rm el}$.  The prediction of $r_Z$ from \nopi is mainly determined
by the deuteron S wave, with a $3.2\%$ correction due to S-D mixing:
\begin{align}
r_{Z,\text{th}}^{D} = 2.691\; \text{fm}-0.086\; \text{fm}=2.605 (91)\; \text{fm},	
\end{align}
with a $3.5\%$ uncertainty from the $N^3$LO correction. The result is consistent with the calculation using the chiral EFT potential~\cite{Nevo:2018hdo}, and agrees with the experimental value
$r_{Z,\text{exp}}=2.593(16)\;\text{fm}$ within $1\sigma$.

The single-nucleon TPE contributions from Eq.~\eqref{eq:rzN} need
inputs for $\tilde{r}_Z^{p,n}$, which accounts for the Zemach, recoil,
and polarizability effects from proton and neutron. The proton TPE
contributions to HFS in $H$ and $\mu$H were determined with high
accuracy by using constraints from HFS spectroscopy
measurements~\cite{Antognini:2022arn,Antognini:2022plb}, while the
neutron TPE effects were determined using a dispersive
calculation~\cite{Tomalak:2019epja,Tomalak:2019prd}. These nucleon TPE
effects are transformed into $\tilde{r}_Z^{p,n}$ for ordinary and
muonic atoms using a scaling approach~\cite{Kalinowski:2018}:
\begin{align}
\tilde{r}_Z^{p,e} =& 0.883(2) \text{ fm}, &\tilde{r}_Z^{p,\mu} =& 0.906(2) \text{ fm},
\nonumber \\
\tilde{r}_Z^{n,e} =& 0.347(38) \text{ fm}, &\tilde{r}_Z^{n,\mu} =& 0.102(39) \text{ fm}.
\end{align}

Table~\ref{tab:tpe-full} summarizes the elastic, polarizability, and
single-nucleon TPE contributions to HFS, comparing our predictions
with experimental values and theoretical results from previous
studies. $E_{\rm el}$ and $E_{\rm pol}$ are consistently calculated in
the \nopi at the same order and are strongly anti-correlated. Therefore, we assign a $3.5\%$ uncertainty to the
combination of $E_{\rm el}+E_{\rm pol}$, reflecting the N$^3$LO
correction. This uncertainty is square summed with the one from the
single-nucleon contributions to estimate the total TPE uncertainty.
In the $^2$H 1S state, the prediction for $E_{\rm TPE}$ is 41.7(2.6)
kHz, which deviates from $\nu_{\rm exp}-\nu_{\rm qed}$~\eqref{eq:tpe-D-exp}
by $7\%$ and falls within $1.3\sigma$ of the combined
theory-experiment uncertainty. For the $\mu^2$H 2S state, the
predicted $E_{\rm TPE}$ is 0.118(9) meV, differing from
$\nu_{\rm exp}-\nu_{\rm qed}$~\eqref{eq:tpe-muD-exp} by $17\%$, within
$1.7\sigma$. $E_{\rm TPE}$ is insensitive to the choice of nucleon form factors. With a different parameterization~\cite{Kelly:2004prc} that suggests a larger proton radius, the predicted $E_{\rm TPE}$ changes by only sub-percent.

\begin{table}[th]
  \centering
    \resizebox{\linewidth}{!}{%
    \begin{tabular}{lL{4.6} L{4.6} L{4.7} L{4.7} }
    \hline\hline
    & \multicolumn{1}{c}{$^2$H (1S)} & \multicolumn{1}{c}{$\mu^2$H (1S)} & \multicolumn{1}{c}{$\mu^2$H (2S)} \\
    \hline
    $E_{\rm el}^{(0)}$ & -41.2 & -1.004 & -0.126 \\
    $E_{\rm el}^{(1)}$ & -1.95 & -0.011 & -0.0014 \\
    $E_{\rm el,sd}^{(0)}$ & 0.97 & 0.030 & 0.0037 \\
    $E_{\rm pol}^{(0)}$ & 122.2 & 3.109 & 0.389 \\
    $E_{\rm pol}^{(1)}$ & -7.8 & -0.129 & -0.016 \\
    $E_{\rm pol,sd}^{(0)}$ & -4.5 & -0.116 & -0.014 \\
    \hline
    $E_{\rm 1p}$~\cite{Antognini:2022arn} & -35.54(8) & -1.018(2) & -0.1272(2) \\
    $E_{\rm 1n}$~\cite{Tomalak:2019epja} & 9.6(1.0)  & 0.08(3) & 0.010(4) \\
    $E_{\rm el}$ & -41.9(1.5) & -0.985(34) & -0.123(4) \\
    $E_{\rm pol}$ & 109.8(3.8) & 2.86(10) & 0.358(13) \\
    \hline
    $E_{\rm TPE}$ & 41.7(2.6) & 0.940(73)  & 0.118(9) \\
    Ref.~\cite{Khriplovich:1995yd,Khriplovich:2003nd} & 43 &       &  \\
    Ref.~\cite{Friar:2005yv,Friar:2005rc}$_{\rm mod}$ & 64.5 &       &  \\
    Ref.~\cite{Kalinowski:2018} &       & 0.304(68) & 0.0383(86) \\
    $\nu_{\rm exp}-\nu_{\rm qed}$~\cite{Eides2001,Krauth:2015nja}   & 45    &      & 0.0966(73)  \\
    \hline\hline
    \end{tabular}}%
  \caption{Single-nucleon, nuclear elastic and polarizability
    contributions to TPE. The subscript `mod' denotes modifications
    made to the findings in Ref.~\cite{Friar:2005rc, Friar:2005yv},
    incorporating nucleon recoil and polarizability effects.}
  \label{tab:tpe-full}%
\end{table}%

In comparison, the TPE effect on HFS in $^2$H was initially calculated
using the zero-range
approximation~\cite{Khriplovich:1995yd,Khriplovich:2003nd}, showing
agreement within $5\%$ with
$\nu_{\rm exp}-\nu_{\rm qed}$~\eqref{eq:tpe-D-exp}. This formalism was
revisited in Ref.~\cite{Faustov:2003pra} to include higher-order
elastic recoil corrections and was extended to estimate polarizability
effects on HFS in $\mu^2$H~\cite{Krauth:2015nja}. This approach
introduces a $33\%$ discrepancy in the deuteron's asymptotic behavior
and an un-quantified model-dependent uncertainty through an arbitrary
energy-integration cutoff. Thus, their agreement with experiments may
be accidental.

Alternatively, the Low-term formalism takes the heavy-nucleon-mass
limit and evaluates
$E_{\rm el}+E_{\rm pol}$ using the closure approximation without
explicitly treating nuclear excitations. However, the approximation
becomes inaccurate when the momentum scale of nuclear excitations is
comparable to $m_l$ or $\gamma$, changing the infrared $q$-dependence
in Eq.~\eqref{eq:H-2g}. Their predicted TPE effect in $^2$H was 46
kHz, whose agreement with
$\nu_{\rm exp}-\nu_{\rm qed}$~\eqref{eq:tpe-D-exp} is accidental due
to the omission of single-nucleon recoil and polarizability
effects. Adding these corrections, the corresponding modified TPE
effects becomes $E_{\rm TPE}^{\rm HFS}(^2\text{H})=64\;\text{kHz}$,
disagreeing with $\nu_{\rm exp}-\nu_{\rm qed}$ by $43\%$. The Low-term
formalism was extended by including higher-order polarizability
corrections to study the TPE effect in
$\mu^2$H~\cite{Kalinowski:2018}, and yielded
$E_{\rm TPE}^{\rm HFS}(\mu^2\text{H})=0.38\;\text{meV}$, accounting
for only 40\% of $\nu_{\rm exp}-\nu_{\rm qed}$~\eqref{eq:tpe-muD-exp}.


\paragraph*{Conclusion.}

The N$^3$LO corrections in \nopi limit the accuracy of our prediction
for the TPE effect on HFS. As another limiting factor, the uncertainty
from the single-nucleon TPE effect may be much larger than anticipated
due to the one-order-of-magnitude discrepancy between the proton
polarizability effects to HFS from
$\chi$PT~\cite{Antognini:2022arn,Hagelstein:2015lph,Hagelstein:2018bdi}
and from dispersion
analysis~\cite{Tomalak:2019epja,Tomalak:2019prd}. This dispute also
raises questions about the predicted neutron polarizability.  A
resolution to the single-nucleon TPE discrepancy requires higher-order
$\chi$PT calculations and future HFS measurements of the 1S state in
$\mu$H~\cite{Sato:2014uza,Pizzolotto:2020fue,Amaro:2021goz}.
Furthermore, to pin down the single-nucleon effects from HFS in
$^2$H and $\mu^2$H, it will be crucial to improve the accuracy of
calculations of the nuclear-structure part of TPE with \nopi beyond
NNLO or with $\chi$EFT, and to measure HFS in $^2$H and $\mu^2$H with
high precision. The formalism developed in this work can also be
applied to future investigations of TPE effects on HFS in other light
atomic systems.


\paragraph*{Acknowledgement.}
We gratefully acknowledge valuable discussions with Daniel Philips,
Sonia Bacca, Thomas Richardson, and Javier Hernandez during the
project. CJ extends gratitude to Yong-Hui Lin for sharing the data on nucleon form factors.
This work was supported by the National Natural Science
Foundation of China (Grant Nos. 12175083, 12335002, and 11805078), the
National Science Foundation (Grant No. PHY-2111426), and the Office of
Nuclear Physics, US Department of Energy (Contract
No. DE-AC05-00OR22725).

\bibliography{tpe-hfs.bib}
\end{document}


\title{Supplemental Material:\\ Nuclear Structure Effects on Hyperfine Splittings in Ordinary and Muonic Deuterium}

\author{Chen Ji}
\affiliation{Key Laboratory of Quark and Lepton Physics, Institute of Particle Physics, Central China Normal University, Wuhan 430079, China}
\affiliation{Southern Center for Nuclear-Science Theory, Institute of Modern Physics, Chinese Academy of Sciences, Huizhou 516000, China}

\author{Xiang Zhang}
\affiliation{Key Laboratory of Quark and Lepton Physics, Institute of Particle Physics, Central China Normal University, Wuhan 430079, China}

\author{Lucas Platter}
\affiliation{Department of Physics and Astronomy, University of
Tennessee, Knoxville, TN 37996, USA}
\affiliation{Physics Division, Oak Ridge National Laboratory, Oak Ridge, TN 37831, USA}


\date{\today}

\begin{abstract}
  This supplemental material includes a brief review of
  pionless effective field theory, specifically focusing on the
  formalism used to compute the two-photon exchange (TPE) process
  discussed in the main text. We provide a detailed presentation of
  the Lagrangian formulation for the $np$ system and its connection to
  both bound and scattering states, and derive the deuteron
  electromagnetic form factors needed for calculating the elastic
  TPE. Furthermore, we outline the formulation of the transition
  matrices and response functions relevant to the inelastic TPE.
\end{abstract}

\maketitle

\section{Pionless effective field theory}
\label{sec:nopi}

\paragraph*{Lagrangian for the  deuteron channel.}
The pionless effective field theory (\nopi) Lagrangian for the
non-relativistic $np$ system in the deuteron channel is composed of
the one-body body kinetic term and two-nucleon spin-triplet contact
interactions~\cite{Chen:1999tn}
\begin{align}
\label{eq:Lnp}
\mathcal{L}=&
N^\dagger\left(i \partial_0+\frac{\nabla^2}{2 m_N}\right) N
-C_0\left(N^T P_i N\right)^\dagger\left(N P_i N\right)
\nonumber \\
& +\frac{1}{8}  C_2 \left[\left(N^T P_i N\right)^\dagger\left(N^T \mathcal{O}^{(2)}_i N\right)+\text { h.c. }\right]
\nonumber \\
& -\frac{1}{16} C_4 \left(N^T \mathcal{O}^{(2)}_i N\right)^\dagger
\left(N^T \mathcal{O}^{(2)}_i N\right)+\cdots~,
\end{align}
where
$\mathcal{O}^{(2)}_i \equiv \overleftarrow{\nabla}^2 P_i -2
\overleftarrow{\nabla} P_i\overrightarrow{\nabla} +P_i
\overrightarrow{\nabla}^2$ is the Galilean invariant operator
involving two derivatives. Here,
$P_i =\frac{1}{\sqrt{8}} \sigma_2 \sigma_i \tau_2$ represents the $np$
spin-triplet projection operator, where $\sigma_i$ and $\tau_i$ are
the spin and isospin Pauli matrices. The operator satisfies
$\text{Tr} [ P_i^\dagger P_j]=\frac{1}{2} \delta_{i j}$. The values of
the low-energy constants (LECs) in Eq.~\eqref{eq:Lnp} are determined
by fitting them to the low-momentum expansion of the $np$ spin-triplet
scattering phase shift
\begin{align}
  \label{eq:ere}
p \cot \delta_t (p) = -\gamma + \frac{\rho}{2}(p^2+\gamma^2) + \cdots,
\end{align}
where the deuteron binding momentum $\gamma=\sqrt{m_N B_d}$ is
directly related to the deuteron binding energy of $B_d=2.2246$
MeV. The effective range for the $np$ spin-triplet state is
given by $\rho=1.764\text{ fm}$. The corresponding LECs are
expanded as
\begin{align}
C_0=&C_{0,-1}+C_{0,0}+C_{0,1}+\cdots~,
\nonumber \\
C_2=&C_{2,-2}+C_{2,-1}+\cdots~,
\nonumber \\
C_4=&C_{4,-3}+\cdots~.
\end{align}
The use of dimensional regularization and power-divergence subtraction
renormalization allows us to express the LECs in terms of the
effective range parameters and the renormalization scale $\mu$
appearing in Eq.~\eqref{eq:ere}. The subscripts in $C_{n,m}$ give the power of momentum $n$
in the operator and $m$ the power of momentum in the coefficient itself
\cite{Chen:1999tn}.
\begin{align}
C_{0,-1}=&-\frac{4 \pi}{m_N} \frac{1}{\mu-\gamma},
& C_{0,0}=&\frac{2 \pi}{m_N} \frac{\rho \gamma^2}{(\mu-\gamma)^2}, 
\nonumber \\
C_{0,1}=&-\frac{\pi}{m_N} \frac{\rho^2 \gamma^4}{(\mu-\gamma)^3},
& C_{2,-2}=&\frac{2 \pi}{m_N} \frac{\rho}{(\mu-\gamma)^2}, 
\nonumber \\
C_{2,-1}=&-\frac{2 \pi}{m_N} \frac{\rho^2 \gamma^2}{(\mu-\gamma)^3}, 
& C_{4,-3}=&-\frac{\pi}{m_N} \frac{\rho^2}{(\mu-\gamma)^3}.
\end{align}
The main text of our paper gives additional operators, such as the S-D
mixing interaction and one- and two-nucleon currents. These terms will
not be repeated here. The $np$ spin-singlet interaction will not be
considered as it does not impact the TPE effect on HFS.

\paragraph*{np  scattering t-matrices.}
The on-shell triplet-state t-matrix for the $np$ scattering state is
expanded at NNLO as follows:
\begin{align}
    \mathcal{A}_t (p,p;E) \approx&
    \left[ \mathcal{A}_t^{(0)} +\mathcal{A}_t^{(1)} +\mathcal{A}_t^{(2)} \right]  (p,p;E)
    \nonumber \\
   & \hspace{-5em} = -\frac{4\pi}{m_N} \frac{1}{\gamma+i p}  \left[ 1+ \frac{\rho}{2} ( \gamma-ip) +  \frac{\rho^2}{4} ( \gamma-ip)^2\right],
\end{align}
where $p=\sqrt{m_N E}$ represents the on-shell momentum, and
$\mathcal{A}_t^{(0)}$, $\mathcal{A}_t^{(1)}$, and
$\mathcal{A}_t^{(2)} $ give the leading order (LO), next-to-leading
order (NLO), and next-to-next-to-leading order (NNLO) contributions to
the on-shell scattering amplitude, respectively.

The half-off-shell $np$ spin-triplet scattering t-matrices at LO, NLO,
and NNLO can be obtained using the following
expressions~\cite{Emmons:2020aov}:
    \begin{align}
    \mathcal{A}_t^{(0)}(k,p;E) =& -\frac{4\pi}{m_N} \frac{1}{\gamma+i p},
    \\
    \mathcal{A}_t^{(1)}(k,p;E) =& - \frac{2\pi}{m_N} \frac{\rho}{\gamma+i p}
  \left[\gamma-i p+\frac{k^2-p^2}{2(\gamma-\mu)}\right],
  \\
    \mathcal{A}_t^{(2)}(k, p; E)
=&-\frac{\pi }{m_N} \frac{\rho^2}{\gamma+i p}
\left[(\gamma-i p)^2  
\right. 
\nonumber \\
&\left.
 + \frac{\gamma-i p}{\gamma-\mu}  \left(1+\frac{\gamma+i p}{\gamma-\mu}\right)
\frac{k^2-p^2}{2}\right],
    \end{align}
where $k$ is the off-shell momentum, which appears in the later sections as the momentum of loop integrals.

 
\paragraph*{Deuteron form factors.}

The deuteron's electric and magnetic form factors are calculated from
matrix elements of single-photon operators between deuteron states.
The deuteron electric charge form factor at NNLO is given by:
\begin{align}
\frac{F_{ed}(q)}{F_{es}(q)} = \frac{1}{1-\gamma \rho } \left[ \frac{4\gamma}{q}\arctan \frac{q}{4\gamma}-\gamma \rho\right]~.
\end{align}
We disregard the relativistic and S-D mixing corrections to the
deuteron's electric charge form factor as they only contribute at
N$^4$LO.

The deuteron's electric quadrupole form factor is generated through
the mixing of S- and D-waves in its ground state. The dominant term
that contributes to it can be expressed in \nopi as:
\begin{align}
\mu_Q \frac{F_{Qd}(q)}{F_{es}(q)} =& -\frac{3\sqrt{2} \eta_{sd}}{q^2 } \left[ 1 -(1+\frac{3q^2}{16\gamma^2}) \frac{4\gamma}{q} \arctan \frac{q}{4\gamma} \right].	
\end{align}
The quadrupole form factor is normalized to $F_{Qd}(0)=1$, and the LO
prediction for the deuteron quadrupole moment is
$\mu_Q = \eta_{sd}/(\sqrt{2}\gamma^2)=0.332\, {\rm
  fm^2}$~\cite{Chen:1999vd,Ando:2004mm}.

The magnetic form factor $F_{md}$ at NNLO in \nopi is derived as~\cite{Chen:1999tn}:
\begin{align}
\frac{F_{md}(q)}{F_{ms}(q)} = &
\left[\frac{2\kappa_0/g_m}{1-\gamma \rho } \left( \frac{4\gamma}{q}\arctan \frac{q}{4\gamma}-1\right)+1\right],
\end{align}
where the inclusion of the two-nucleon magnetic current is crucial for
accurately reproducing the measured deuteron magnetic anomalous
factor.

\section{Transition matrices in virtual photon excitation}
\label{sec:dgnp}
\begin{figure}[htb]
\begin{center}
\includegraphics[width=0.8\linewidth]{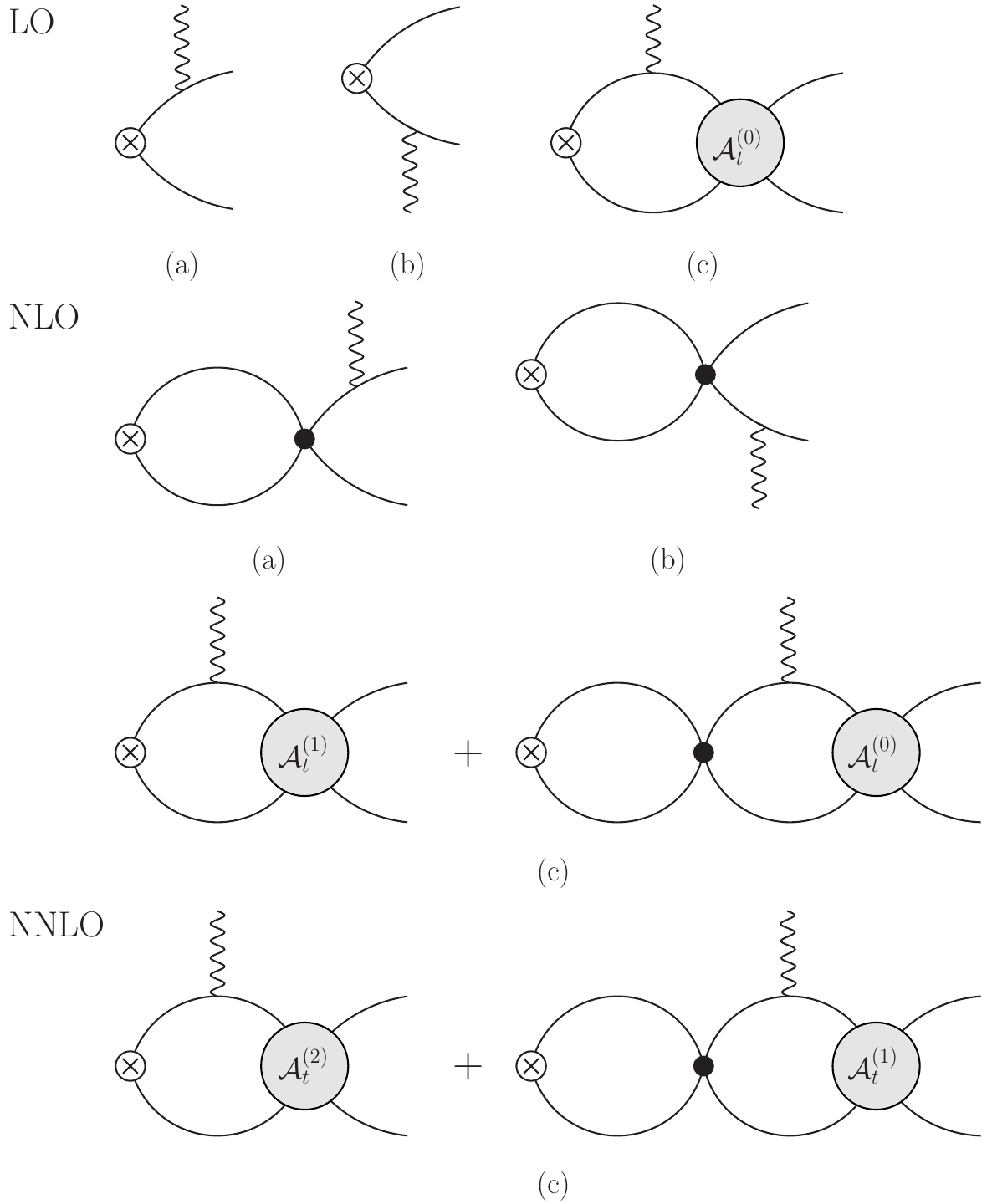}
\caption{ Deuteron transition matrices at different orders in \nopi.
  Group (a) and (b) include diagrams that involve plane-wave excited
  states with a $NN$ permutation. Group (c) consists of diagrams
  involving $NN$ final-state interactions.  }
 \label{fig:diag-full}
\end{center}
\end{figure}

The deuteron's transition matrix elements from virtual photon
excitation are obtained from the Feynman diagrams depicted in
Fig.~\ref{fig:diag-full}. In the \nopi expansion, contributions to the
transition matrices can be classified into three groups referred to as
(a), (b), and (c). The photon-nucleon vertex encompass the
charge-density, convection-current, and magnetic-current operators.

\paragraph*{Charge density transition.}

The transition matrices resulting from the charge density operator can
be determined by employing the formulas given in
Ref.~\cite{Emmons:2020aov}:
\begin{align}
    \mathcal{T}_a
    =& |Z_d|^{\frac{1}{2}} i \tilde{T}_a N^T(\vec{u}) 
       \Pi_+ \left[F_{e s}(q)+ \tau_3 F_{e v}(q)\right]  N(\vec{v}),
\\
\mathcal{T}_b
    =& |Z_d|^{\frac{1}{2}} i \tilde{T}_b
    N^T(\vec{u}) \left[F_{e s}(q) + \tau_3 F_{e v}(q)\right] 
   \Pi_+  N(\vec{v}),
\\
\mathcal{T}_c
     =&  2 |Z_d|^{\frac{1}{2}}   i \tilde{T}_c  F_{es}(q) 
      N^T(\vec{u})  \Pi_+  N(\vec{v}), 
\end{align}
where the momenta of the two outgoing nucleon momenta are
$\vec{u}=-\vec{p}+\frac{\vec{q}}{2}$ and
$\vec{v}=\vec{p}+\frac{\vec{q}}{2}$, and $\vec{q}$ is the photon
momentum. $Z_d = -8\pi\gamma/[m_N^2 (1-\rho \gamma)]$ is the
renormalization strength of the deuteron field.  We define
spin-triplet projection operators with spherical indices that enable
to use to carry out calculations for matrix elements with a specific
$I_z$-projection
\begin{align}
\Pi_{\pm}=& i\left(P_2\pm i P_1\right) /\sqrt{2},\quad \Pi_0 = P_3~.
\end{align}
These satisfy
${\rm Tr} \left[\Pi_\alpha \Pi_\beta^\dagger\right]=\frac{1}{2}
\delta_{\alpha \beta}$, where $ \alpha$ and $\beta$ take the values
$-$, $0$, or $+$.

Summing the transition matrices through NNLO, we obtain the amplitudes
$\tilde{T}_{a,b,c}$:
\begin{align}
\tilde{T}_a =& - \frac{m_N}{ \gamma^2 + \left(\vec{p}-\frac{\vec{q}}{2}\right)^2 } + \frac{ m_N \rho}{4(\mu-\gamma)}~,
\\
\tilde{T}_b =& - \frac{m_N}{ \gamma^2 + \left(\vec{p}+\frac{\vec{q}}{2}\right)^2 } + \frac{ m_N \rho}{4(\mu-\gamma)}~,
\\
\tilde{T}_c 
    =&
    \mathcal{A}_t^{(0)} (E_f ) \left[ \mathcal{J}_0 + \frac{\rho}{2}\left((\gamma-ip)\mathcal{J}_0+ \frac{m_N^2}{4 \pi} \right)
    \right.
    \nonumber \\
    &\left.
    \left(1+\frac{\rho}{2}(\gamma-ip)\right) \right] - \frac{ m_N \rho}{4(\mu-\gamma)}~,
\end{align}
where the loop integral $\mathcal{J}_0$ is defined in Eq.~\eqref{eq:J0}.


\paragraph*{Magnetic current transition.}
By using the one-body magnetic current operator in the photon-nucleon
vertices shown in Fig.~\ref{fig:diag-full}, we obtain the transition
matrices induced by the magnetic current:
\begin{align}
\vec{\mathcal{M}}_a
=&  \frac{1}{m_N} |Z_d|^{\frac{1}{2}} i \tilde{M}_a
    N^T(\vec{u})
    \Pi_+ 
    i (\vec{\sigma} \times \vec{q}) 
    \nonumber \\
    &
    \left[\kappa_0 F_{m s}(q)+\tau_3 \kappa_1 F_{m v}(q)\right] 
   N(\vec{v}) ,
    \\
\vec{\mathcal{M}}_b
=& \frac{1}{m_N}|Z_d|^{\frac{1}{2}}  i \tilde{M}_b
    N^T(\vec{u})  
    i (\vec{\sigma}^T \times \vec{q}) 
    \nonumber \\
    & \left[\kappa_0 F_{m s}(q)+\tau_3 \kappa_1 F_{m v}(q)\right] 
    \Pi_+ N(\vec{v}) ,
   \\
\vec{\mathcal{M}}_c
=& \frac{4}{ m_N}|Z_d|^{\frac{1}{2}} i \tilde{M}_c \text{Tr}\left[\Pi_+ i( \vec{\sigma} \times \vec{q}) \Pi_\alpha^\dagger \right] 
\nonumber \\
 & 
\kappa_0 F_{m s} (q) N^T(\vec{u}) \Pi_\alpha  N(\vec{v}).
\end{align}

\begin{figure}[htb]
\begin{center}
\includegraphics[width=0.8\linewidth]{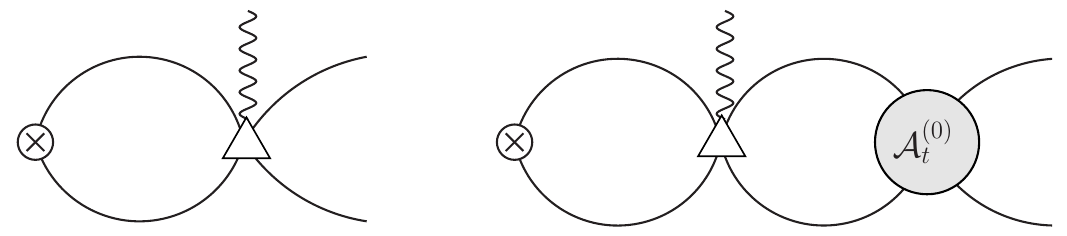}
\caption{ Th deuteron transition matrices with the inclusion of the
  two-nucleon current, depicted by the upward triangle.  }
 \label{fig:diag-L2}
\end{center}
\end{figure}

The two-nucleon current shown in Fig.~\ref{fig:diag-L2} contributes to
the magnetic transition matrices at NNLO. I t can be incorporated into
the amplitude $\tilde{M}_c$.  By summing up contributions to the
magnetic transition through NNLO, we obtain the amplitudes
$\tilde{M}_{a,b,c}$ as
\begin{align}
\tilde{M}_a = \tilde{T}_a; \quad
\tilde{M}_b = \tilde{T}_b; \quad
\tilde{M}_c =    
\tilde{T}_c
+ \frac{m_N(\frac{g_m}{\kappa_0}-2)}{4\gamma (\gamma+ip)}.
\end{align}

\paragraph*{Convection current transition.}
The transition matrices induced by the convection current can be
obtained from the diagrams shown in Fig.~\ref{fig:diag-full} with the
appropriate single-photon operator used in the calculation. We obtain
\begin{align}
    \vec{\mathcal{C}}_a
    =& |Z_d|^{\frac{1}{2}}  \frac{\vec{p}}{m_N}  i \tilde{C}_a
    N^T(\vec{u}) 
     \Pi_+ 
    \left[F_{e s}(q)+\tau_3 F_{e v}(q)\right] 
    N(\vec{v})~,
    \\
    \vec{\mathcal{C}}_b
    =&- |Z_d|^{\frac{1}{2}}  \frac{\vec{p}}{m_N} i \tilde{C}_b    N^T(\vec{u}) 
     \left[F_{e s}(q)+ \tau_3 F_{e v}(q)\right]  
  \Pi_+ N(\vec{v})~,
    \\
    \vec{\mathcal{C}}_c
     =& 2 |Z_d|^{\frac{1}{2}} \frac{\vec{q}}{m_N} i \tilde{C}_c^{(0)} F_{es}(q) 
    N^T(\vec{u}) \Pi_+    N(\vec{v})~. 
\end{align}
The two-nucleon convection current gives an additional NLO
contribution to the transition matrices, as depicted in
Fig.~\ref{fig:diag-L2}. Evaluation of the diagram leads to
\begin{align}
 \vec{\mathcal{C}}_{d}
     = |Z_d|^{\frac{1}{2}} \frac{\vec{p}}{m_N}   i \tilde{C}_{d}
    F_{ev}(q) 
    N^T(\vec{u}) \Pi_+ \tau_3   N(\vec{v}).
\end{align}
Since the emergence of the convection current is already suppressed by
$1/m_N$ in TPE, we calculate the convection transition matrices up to
NLO. This allows us to obtain the corresponding amplitudes, denoted as
$\tilde{C}_{a,b,c,d}$:
\begin{align}
 \tilde{C}_a  =& \tilde{T}_a;\quad
 \tilde{C}_b = \tilde{T}_b; \quad
 \tilde{C}_{d} = -\frac{m_N \rho}{2(\mu-\gamma)};
\nonumber \\
 \tilde{C}_c 
=& \frac{m_N\omega}{q^2} \mathcal{A}_t^{(0)}(p,p;E_f) \left\{ \mathcal{J}_0 + \frac{m_N}{4\pi\omega}(\gamma+ip) \right.
\nonumber \\
& \left.+ \frac{\rho}{2}\left[ (\gamma-ip) \mathcal{J}_0 + \frac{m_N^2}{4\pi}\right]\right\}.
\end{align}


\paragraph*{S-D mixing in transition.}
\label{sec:sd-mixing}

\begin{figure}[th]
\begin{center}
\includegraphics[width=0.8\linewidth]{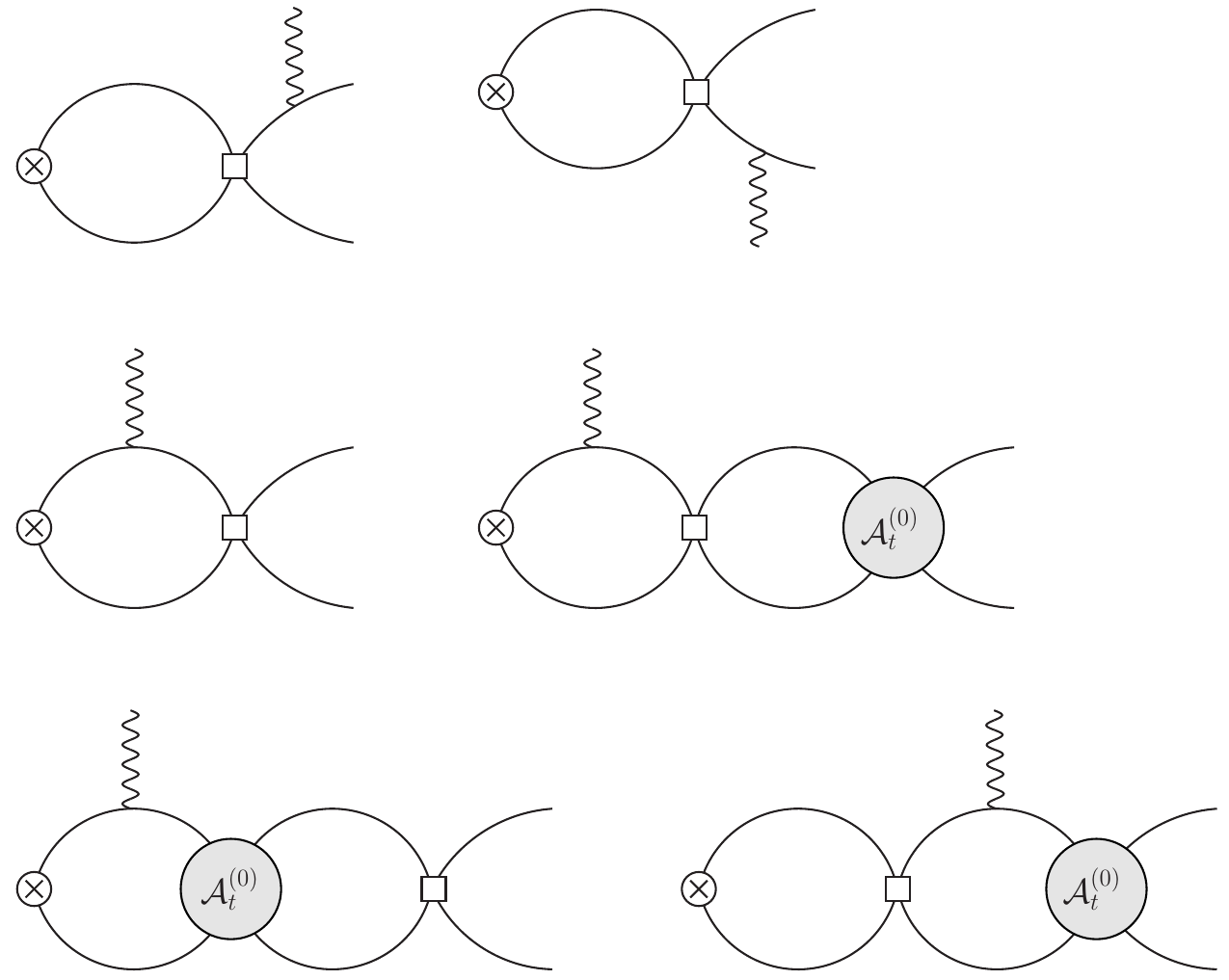}
\caption{
Deuteron transition matrices incorporating the S-D mixing operator, represented by the square.
}
 \label{fig:diag-SD}
\end{center}
\end{figure}

Since the S-D mixing operator emerges at NNLO, we only evaluate its dominant correction to the response function $S^{(0)}$. Its contribution to the transition matrices, shown in Fig.~\ref{fig:diag-SD}, is as follows:
\begin{align}
  \mathcal{Q}_{a}=&
2 |Z_d|^{\frac{1}{2}} i\tilde{Q}_a \text{Tr}[\Pi_+ P_i^\dagger] N^T(\vec{u}) P_j  
\nonumber \\
& \left[ F_{e s}(q)+ \tau_3 F_{e v}(q)\right] N(\vec{v})
(u_i u_j-\frac{1}{3}u^2\delta_{i j} ),
\\
 \mathcal{Q}_{b}=&
2 |Z_d|^{\frac{1}{2}} i\tilde{Q}_b \text{Tr}[\Pi_+ P_i^\dagger] N^T(\vec{u}) 
\nonumber \\
& \left[ F_{e s}(q)+ \tau_3 F_{e v}(q)\right] P_j  N(\vec{v})
(v_i v_j -\frac{1}{3}v^2 \delta_{i j}),
\\
\mathcal{Q}_{c} =&  4|Z_d|^{\frac{1}{2}} i \tilde{Q}_c  F_{e s}(q) 
    \text{Tr}[\Pi_+ P_i^\dagger] N^T(\vec{u}) P_j  N(\vec{v})
    \nonumber \\
    & (q_i q_j - \frac{1}{3} q^2 \delta_{ij}),
    \\
\mathcal{Q}_{d} =&  4|Z_d|^{\frac{1}{2}} i \tilde{Q}_d  F_{e s}(q) 
    \text{Tr}[\Pi_+ P_i^\dagger] N^T(\vec{u}) P_j  N(\vec{v})
    \nonumber \\
    & ( p_i p_j -\frac{1}{3} p^2 \delta_{ij}),
\end{align}
where the amplitudes $\tilde{Q}_{a,b,c,d}$ are given by
\begin{align}
\tilde{Q}_a
=& - \frac{3 \eta_{sd} m_N }{\sqrt{2}\gamma^2 (\gamma^2+\vec{u}^2 )} ;\;
  \tilde{Q}_d 
  = -\frac{6\sqrt{2}\pi \eta_{sd}\mathcal{J}_0 }{m_N \gamma^2  (\gamma+ip)} ;
\\
\tilde{Q}_b
=& - \frac{3 \eta_{sd} m_N }{\sqrt{2}\gamma^2 ( \gamma^2+\vec{v}^2 ) };\;
\tilde{Q}_c = -\frac{6\sqrt{2}\pi \eta_{sd} \bar{J}_Q }{m_N \gamma^2 q^2 (\gamma+ip)}~.
\end{align}
with $\bar{J}_Q$ defined in Eq.~\eqref{eq:JQ}.


\paragraph*{Response functions.}
The transition matrices are used in the response functions discussed
in the main text. After trace evaluation (see Eqs.~\eqref{eq:traces1}, \eqref{eq:traces2}, and \eqref{eq:traces3})
and integrating over the solid angles of $\hat{p}$ and $\hat{q}$, we
derive the response functions $S^{(0)}$, its S-D mixing correction
$S^{(0)}_{\rm sd}$, and $S^{(1)}$ as
\begin{align}
\label{eq:S0-an}
S^{(0)}(\omega,q) 
=& \frac{p|Z_d|}{12\pi^2}  {\rm Re} \left[ U_{mt,s}^{[0]} \kappa_0 F_{m s} F_{e s} + U_{mt,v}^{[0]} \kappa_1 F_{m v} F_{e v} \right],	
\\
\label{eq:S1-an}
S^{(1)}(\omega,q) 
=& \frac{p|Z_d|}{12\pi^2m_N} \text{Re}\left[( qp U_{mc,s}^{[1]} + q^2 W_{mc,s}^{[0]}) \kappa_0 F_{ms} F_{es}
\right.
\nonumber \\
& \left.  + qp U_{mc,v}^{[1]} \kappa_1 F_{mv} F_{ev} \right],
\\
\label{eq:S2-an}
S^{(0)}_{\rm sd}(\omega,q) =& \frac{p|Z_d|}{36\pi^2} \text{Re} \left[ 
( \frac{q^2}{4} V_{sd,s}^{[0]} - p q X_{sd,s}^{[1]} + p^2 W_{sd,s}^{[2]}  ) \kappa_0 
\right.
\nonumber \\
& \hspace{-8ex} \left. F_{es}F_{ms}
+ 
( \frac{q^2}{4} V_{sd,v}^{[0]} - p q X_{sd,v}^{[1]} + p^2 V_{sd,v}^{[2]} ) \kappa_1 F_{ev}F_{mv}
\right].
\end{align}
where the kernel functions $U_{mt,s/v}$, $U_{mc,s/v}$, $W_{mc,s}$, $V_{sd,s/v}$, $X_{sd,s/v}$, $W_{sd,s}$ are given by
\begin{align}
  U_{m t, s} =& (\tilde{M}_a+\tilde{M}_b+2 \tilde{M}_c)(\tilde{T}_a+\tilde{T}_b+2 \tilde{T}_c)^*~,
  \\
  U_{m t, v} =& (\tilde{M}_a-\tilde{M}_b)(\tilde{T}_a-\tilde{T}_b)^*~,
  \\
  U_{mc, s}=&(\tilde{M}_a+\tilde{M}_b+2 \tilde{M}_c)(\tilde{C}_a-\tilde{C}_b)^*~,
  \\
  W_{mc, s}=&2(\tilde{M}_a+\tilde{M}_b+2 \tilde{M}_c) \tilde{C}_c^*~,
  \\
  U_{m c, v}=&(\tilde{M}_a-\tilde{M}_b)(\tilde{C}_a+\tilde{C}_b+\tilde{C}_d)^*~,
  \\
  V_{sd,s} =& \left(\tilde{M}_a+\tilde{M}_b+2 \tilde{M}_c\right)\left(\tilde{Q}_a+\tilde{Q}_b+8 \tilde{Q}_c\right)^*~,
  \\
  W_{sd,s} =&  \left(\tilde{M}_a+\tilde{M}_b+2 \tilde{M}_c\right)\left(\tilde{Q}_a+\tilde{Q}_b+2 \tilde{Q}_d\right)^*~,
  \\
  X_{sd,s} =& \left(\tilde{M}_a+\tilde{M}_b+2 \tilde{M}_c\right)\left(\tilde{Q}_a-\tilde{Q}_b\right)^*~,
  \\
  V_{sd,v} =& \left(\tilde{M}_a-\tilde{M}_b\right)\left(\tilde{Q}_a-\tilde{Q}_b\right)^*~,
  \\
  X_{sd,v} = &  \left(\tilde{M}_a-\tilde{M}_b\right)\left(\tilde{Q}_a+\tilde{Q}_b\right)^*~.
\end{align}
The superscript $[n]$ in Eqs.~(\ref{eq:S0-an}-\ref{eq:S2-an})
represents the $n$th moment of the Legendre expansion of the kernel
function. For instance,
\begin{align}
  U_{mt,s}^{[n]} = \frac{1}{2} \int_{-1}^1 U_{m t, s} P_n (\hat{p}\cdot\hat{q}) d (\hat{p}\cdot\hat{q})~.
\end{align}


\appendix

\section{Relevant loop integrals}
\label{app:loop}
The three-point loop integral that involves the insertion of the charge or
magnetic operator is given by
\begin{align}
\label{eq:J0}
    \mathcal{J}_0
    =&\int \frac{d^4 l}{(2 \pi)^4}  iS \left(-B_d+l_0, \vec{l}\right) iS\left(-B_d+l_0+\omega, \vec{l}+\vec{q}\right)
    \nonumber \\
& \qquad\times iS \left(-l_0, \vec{l}\right)
\nonumber \\
=& -m_N^2 \int  \frac{d^3 l}{(2 \pi)^3}   \frac{1}{\left(l^2+\gamma^2\right)\left[\left(\vec{l}+\frac{\vec{q}}{2}\right)^2-p^2 -i\epsilon\right]}
\nonumber \\
=&-\frac{m_N^2}{2 \pi q}  \arctan \frac{q / 2}{\gamma-i p}~.
\end{align}
The loop integral $\vec{\mathcal{J}}_{c,0}$ results from the
convection current operator inserted in the diagrams shown
Fig.~\ref{fig:diag-full}. It contributes to the LO part of the
amplitude $\tilde{C}_c$ and is evaluated as follows:
\begin{align}
    \label{eq:Jc0}
\vec{\mathcal{J}}_{c,0}
= &\int \frac{d^4 l }{(2 \pi)^4}
i S \left(-B_d+l_0, \vec{l}\right) 
iS\left(-B_d+l_0+\omega, \vec{l}+\vec{q}\right)
\nonumber \\
& iS \left(-l_0, \vec{l}\right) 
\frac{2\vec{l}+\vec{q}}{2 m_N}
\nonumber \\
=& -m_N \int \frac{d^3 l}{(2 \pi)^3} \frac{\vec{l}+\frac{\vec{q}}{2}}{\left(l^2+\gamma^2\right)\left[\left(\vec{l}+\frac{\vec{q}}{2}\right)^2-\vec{p}^2-i\epsilon\right]}
\nonumber \\
=&\frac{\vec{q}}{q^2}\left[\frac{m_N}{4 \pi}(\gamma+i p)+\omega \mathcal{J}_0\right]~,
\end{align}
To evaluate the NLO component of $\tilde{C}_c$, we need the loop
integral with the convection current inserted in
Fig.~\ref{fig:diag-full}. This integral is divided into two parts:
$\vec{\mathcal{J}}_{c,2}=\vec{\mathcal{J}}_{c,2}^{(0)}+\vec{\mathcal{J}}_{c,2}^{(1)}$,
where $\vec{\mathcal{J}}_{c,2}^{(0)}$ includes the LO final-state
interaction, and gives a zero contribution.
\begin{align}
\vec{\mathcal{J}}_{c,2}^{(0)} 
=&  
\frac{C_{2,-2}}{2} \mathcal{I}_0\left(-B_d\right) 
\int \frac{d^4 l}{(2 \pi)^4}  iS\left(-B_d+l_0+\omega, \vec{l}+\vec{q}\right) 
\nonumber \\
&iS\left(-l_0, -\vec{l}\right) i S\left(-B_d+l_0, \vec{l}\right) \frac{2\vec{l}+\vec{q} }{2 m_N}\left(l^2+\gamma^2\right)
\nonumber \\
=&-\frac{C_{2,-2}}{2} \mathcal{I}_0\left(-B_d\right) 
m_N \int \frac{d^3 l}{(2 \pi)^3} \frac{\vec{l}+\frac{\hat{q}}{2}}{\left(\vec{l}+\frac{\vec{q}}{2}\right)^2-p^2-i \epsilon}
\nonumber \\
=&0~.
\end{align}
$\vec{\mathcal{J}}_{c,2}^{(1)}$ involves the NLO final-state interaction, and is given by
\begin{align}
\vec{\mathcal{J}}_{c,2}^{(1)} 
=& \int \frac{d^4 l}{(2 \pi)^4} i S\left(-B_d+l_0, \vec{l}\right) iS\left(-B_d+l_0+\omega, \vec{l}+\vec{q}\right) 
\nonumber \\
& iS\left(-l_0, -\vec{l}\right)
\frac{ \mathcal{A}_t^{(1)}\left(\vec{l}+\frac{\vec{q}}{2}, \vec{p}, E_f \right) }{ \mathcal{A}_t^{(0)}\left(E_f\right) }
\frac{2\vec{l}+\vec{q}}{2m_N}
\nonumber \\
=& \frac{\rho}{2}\left[
(\gamma-i p) \vec{\mathcal{J}}_{c,0}-\frac{m_N}{2(\gamma-\mu)} \int \frac{d^3 l}{(2 \pi)^3} \frac{\vec{l}+\frac{\vec{q}}{2}}{l^2+\gamma^2}
\right]
\nonumber \\
=& \frac{\rho \omega \vec{q}}{2q^2}
\left[\frac{m_N^2}{4 \pi }+(\gamma-i p) \mathcal{J}_0\right].
\end{align}

The transition amplitude that involves the S-D mixing operator, shown
in Fig.~\ref{fig:diag-SD}, contains two three-point loop integrals.
\begin{align}
 \mathcal{J}_{Q1}=&\int \frac{d^{4} l}{(2 \pi)^{4}}  iS\left(l_0+\omega, \vec{l}+\vec{q}\right) iS\left(-l_0-B_d,-\vec{l}\right)
 \nonumber \\
& i S\left(l_0, \vec{l}\right)
\left[\left(l_i+\frac{q_i}{2}\right) \left(l_j+\frac{q_j}{2}\right)-\frac{1}{3}\left(\vec{l}+\frac{\vec{q}}{2}\right)^2 \delta_{i j}\right]
 \nonumber \\
 =&-m_N^2 \int \frac{d^3 l}{(2 \pi)^3} \frac{l_i l_j-\frac{1}{3} l^2 \delta_{i j}}{\left[\left(\vec{l}-\frac{\vec{q}}{2}\right)^2+\gamma^2\right]\left[l^2-p^2-i \epsilon\right]}
 \nonumber \\
=&\bar{J}_{Q1}\left(q_i q_j-\frac{1}{3} q^2\delta_{i j}\right)/q^2~,
\\
\mathcal{J}_{Q2} =&
\int \frac{d^{4} l}{(2 \pi)^{4}} iS\left(l_0+\omega, \vec{l}+\vec{q}\right) iS\left(-l_0-B_d,-\vec{l}\right)~
\nonumber \\
&\qquad\times i S\left(l_0, \vec{l}\right) \left[l_i l_j-\frac{1}{3}\vec{l}^2 \delta_{i j}\right]~,
\nonumber \\
=&\bar{J}_{Q2}\left(q_i q_j-\frac{1}{3} q^2\delta_{i j}\right)/q^2~,
\end{align}
where the amplitudes $\tilde{J}_{Q1}$ and $\tilde{J}_{Q2}$ are given as
\begin{align}
    \bar{J}_{Q1}=& \frac{m_N^2}{16 \pi q^3} 
    \left\{  q\left[q^2 \gamma+6 m_N \omega(\gamma+i p)\right]  \right.
    \nonumber \\
    & \left. +4\left[q^2 p^2-3 m_N^2 \omega^2\right] \arctan \frac{q / 2}{\gamma-i p}\right\}~,
\\
    \bar{J}_{Q2}=& -\frac{m_N^2}{16 \pi q^3} 
    \left\{  q\left[ i q^2 p +6 m_N \Omega(\gamma+i p)\right] 
    \right. 
    \nonumber \\
    &\left.
    +4\left[q^2 \gamma^2 +3 m_N^2 \Omega^2\right] \arctan \frac{q / 2}{\gamma-i p}\right\}~,
\end{align}
with $\Omega= (-p^2-\gamma^2+q^2/4)/m_N$. Their combination yields 
\begin{align}
\label{eq:JQ}
\bar{J}_Q 
=& \bar{J}_{Q1} + \bar{J}_{Q2} 
\nonumber \\
=& \frac{m_N^2}{32\pi q^3} \left\{ 
2 q(\gamma-i p) \left[ q^2 +12(\gamma+ip)^2\right]
- \left[48(p^2+\gamma^2)^2
\right. \right.
\nonumber \\
& \left.\left.
+8q^2(\gamma^2-p^2)+3q^4\right] \arctan \frac{q/2}{\gamma-ip}
\right\}~.
\end{align}

\section{Trace algebra summation}
\label{app:trace}
The $np$ spin-isospin projection operators, denoted as $\Pi_\alpha$,
satisfy the trace identity
$\text{Tr}\left[\Pi_\alpha \Pi_\beta^\dagger\right]=\frac{1}{2}
\delta_{\alpha \beta}$ and
$\text{Tr}\left[\Pi_\alpha \tau_3 \Pi_\beta^\dagger\right] = 0$, with
$\alpha,\beta=0,\pm$.  These projection operators anti-commute with
the isospin operator $\tau_3$.

The nuclear response functions are connected to the transition
matrices by employing the trace summation:
\begin{align}
  \label{eq:traces1}
&\text{Tr}\left[\Pi_+ \vec{\sigma} \Pi_+^\dagger\right] = \text{Tr} \left[\vec{\sigma}^T \Pi_+  \Pi_+^\dagger\right] = (0,0,\frac{1}{2}),
  \\
    \label{eq:traces2}
&\text{Tr}\left[\Pi_+ \vec{\sigma} \tau_3 \Pi_+^\dagger\right] = \text{Tr} \left[\tau_3 \vec{\sigma}^T \Pi_+  \Pi_+^\dagger\right] = (0,0,0),
  \\
    \label{eq:traces3}
&\text{Tr}\left[\Pi_+ P_i^\dagger\right]^* \text{Tr}\left[\Pi_+ \vec{\sigma} P_j^\dagger\right] = \vec{b}_{ij},
\end{align}
where $\vec{b}_{ij}$ equals
\begin{align}
\vec{b}_{ij} 
=\frac{1}{8}\left\{\pcolvec{0 & 0 & -1 \\0 & 0 & -i \\0 & 0 & 0}_{ij},\pcolvec{0 & 0 & i \\0 & 0 & -1 \\0 & 0 & 0}_{ij},\pcolvec{1 & -i & 0 \\i & 1 & 0 \\0 & 0 & 0}_{ij}\right\}.
\end{align}

\bibliography{tpe-hfs.bib}